\documentclass{ws-procs9x6}

\begin{document}
\title{Towards Precision Measurements of $\gamma$: CLEO-c's Pivotal Role}
\author{ANDREW~S. POWELL}
\address{University of Oxford, Denys Wilkinson Building, Oxford, OX1 3RH, UK}

\maketitle

\abstracts{
Strategies that utilise the interference effects within $\rm B^{\pm}\!\to DK^{\pm}$ decays hold great potential for improving our 
sensitivity to the CKM angle $\gamma$. However, in order to exploit fully this potential, detailed knowledge of the D meson decay 
structure is required. This essential information can be obtained from the quantum correlated $\psi(3770)$ datasets at CLEO-c. Results 
of such analyses involving the decay mode $\rm D \to K\pi\pi\pi$, and their importance in the context of LHCb, will be presented.}

\section{Introduction}
A means of testing the internal consistency of the Cabbibo-Kaboyashi-Maskawa (CKM) model, whilst simultaneously searching for 
signatures of New Physics, is to perform precision measurements of the angles that compose the unitarity triangle: $\alpha$, $\beta$ 
and $\gamma$. While $\beta$ has been measured with extremely high precision at the B-factories ($20.5 \pm 
1.0^{\circ}$)\cite{CKMFitter}, determination of the other two angles currently presents a considerable experimental challenge; most 
notably $\gamma$ which is only constrained by direct measurements with a precision of $\sim \pm 30^{\circ}$\cite{CKMFitter}. One of 
the most promising ways of determining the angle $\gamma$ is through strategies that exploit the interference within $\rm B^{\pm}\!\to 
DK^{\pm}$ decays.\footnote{Here and subsequently, D will denote a $\rm D^{0}$ or $\rm \bar{D}^{0}$} The most straightforward of these 
strategies considers two-body final states of the D meson, however, an abundance of additional information can be gained from 
strategies that consider multi-body final states instead. In order to exploit fully the wealth of statistics soon to arrive at the 
LHC, the LHCb\cite{LHCbTP,LHCbTDR} experiment plans to utilise all such strategies in its analysis. Multi-body strategies, however, 
only have significant sensitivity when combined with detailed knowledge of the D meson decay structure. Fortunately, the parameters 
associated with the specific multi-body final states needed for these analyses can be extracted from correlations within 
CLEO-c\cite{CLEO} $\psi(3770)$ data. 

\section{Determination of the CKM angle $\gamma$ from $\rm B^{\pm}\!\to DK^{\pm}$}\label{sec:BDK}
The interference between $\mathrm{B}^{-}\!\to \mathrm{D}^{0}\mathrm{K}^{-}$ and $\mathrm{B}^{-}\!\to 
\mathrm{\bar{D}}^{0}\mathrm{K}^{-}$ decays, and equally between their CP conjugate states, provides a clean mechanism for the 
extraction of the CKM angle $\gamma$ when the $\mathrm{D}^{0}$ and $\mathrm{\bar{D}}^{0}$ mesons decay to a common final state, 
$f_{\mathrm{D}}$. Decay rates in these channels have the following amplitude ratio
\begin{equation}
\frac{
A(\mathrm{B}^{-}\!\rightarrow \mathrm{\bar{D}}^{0} \mathrm{K}^{-})}
{A(\mathrm{B}^{-}\!\rightarrow \mathrm{D}^{0} \mathrm{K}^{-})} = r_{B}e^{i(\delta_{B} - \gamma)},
\end{equation}
\noindent
which is a function of three quantities: the ratio of the amplitudes absolute magnitudes $r_{B}$, a CP invariant strong phase 
difference $\delta_{B}$, and the CKM weak phase $\gamma$. Generically, the amplitude for the complete decay  $\mathrm{B}^{-}\! 
\rightarrow \mathrm{D}(f_{\mathrm{D}})\mathrm{K}^{-}$, normalised to the favoured $\mathrm{B}\!\rightarrow \mathrm{DK}$ amplitude, is 
defined as
\begin{equation}
\frac{\displaystyle
A(\mathrm{B}^{-}\!\rightarrow \mathrm{D}(f_{\mathrm{D}})\mathrm{K}^{-})}{A(\mathrm{B}^{-}\!\rightarrow \mathrm{D^{0}}\mathrm{K}^{-})} 
=  A_{{D}^{0}} + r_{B}e^{i(\delta_{B} - \gamma)}A_{\bar{D}^{0}}, 
\end{equation}
where $A_{{D}^{0}}$ and $A_{\bar{D}^{0}}$ represent the amplitudes for the $\mathrm{D^{0}}$ and $\mathrm{\bar{D}^{0}}$ decays, 
respectively. Due to colour suppression $r_{B} < 0.13$ @ $90\%$ CL\cite{CKMFitter}; therefore, the interference is generally small. A 
variety of strategies exist, however, that attempt to resolve this and maximise the achievable sensitivity to $\gamma$. One such 
tactic is to consider multi-body final states of the D meson.

\subsection{ADS Formalism}
Atwood, Dunietz and Soni (ADS)\cite{ADS} have suggested considering D decays to non-CP eigenstates as a way of maximising sensitivity 
to $\gamma$. Final states such as $\rm K^{-}\pi^{+}$, which may arise from either a Cabibbo favoured $\rm D^{0}$ decay or a doubly 
Cabibbo suppressed $\rm \bar{D}^{0}$ decay, can lead to large interference effects and hence provide particular sensitivity to 
$\gamma$. This can be observed by considering the rates for the two possible $\rm B^{-}$ processes:
\begin{eqnarray}
\Gamma ({\rm B^{-}\!\to (K^{-}\pi^{+})_{D}K^{-}}) & \propto & 1 \, + \, (r_B \, r_D^{\rm K\pi})^2 \, \nonumber \\
{} & {} & + \, 2 \, r_B \, r_D^{\rm K\pi} \, \cos \left( \delta_B \, - \, \delta_D^{\rm K \pi} \, - \, \gamma \right),
\label{eq:fav1} \\
\Gamma ({\rm B^{-}\!\to (K^{+}\pi^{-})_{D}K^{-}}) & \propto & r_B^2 \, + \, {(r_D^{\rm K\pi})}^2 \, \nonumber \\
{} & {} & + \, 2 \, r_B \, r_D^{\rm K\pi} \, \cos \left( \delta_B \, + \, \delta_D^{\rm K \pi} \, - \, \gamma \right), 
\label{eq:dis1}
\end{eqnarray}
\noindent where $r_{D}^{\rm K\pi}$, [$(61.3 \pm 0.7) \times 10^{-3}$]\cite{PDG}, parameterises the relative suppression between 
$A_{{D}^{0}}$ and $A_{\bar{D}^{0}}$, and $\delta_D^{\rm K \pi}$, [$(22^{+14}_{-15})^{\circ}$]\cite{TQCA}, the relative strong phase 
difference.

Since $r_{B}$ and $r_{D}^{\rm K\pi}$ are expected to be similar in magnitude, it can be seen that whilst Eq.~(\ref{eq:dis1}) is the 
more suppressed of the two rates, it provides greater sensitivity to $\gamma$ as a result of the interference term appearing at 
leading order. Through considering the other two rates associated with the $\rm B^{+}$ decay, and combining this with information from 
decays to the CP-eigenstates $\rm K^{+}K^{-}$ and $\pi^{+}\pi^{-}$, an unambiguous determination of $\gamma$ can be made. The expected 
one-year sensitivity to $\gamma$ from these six rates is estimated to be 8-$10^{\circ}$ at LHCb\cite{LHCbADS}, depending on the value 
of $\delta_D^{\rm K \pi}$.

\subsection{Multi-body Extension to the ADS Method}
A wealth of additional statistics can be gained from considering multi-body decays of the D meson. In the case of the ADS method, 
these are states involving a charged kaon and some ensemble of pions, such as $\rm D \to K^{-}\pi^{+}\pi^{0}$ and $\rm D \to 
K^{-}\pi^{+}\pi^{-}\pi^{+}$. However, a complication comes from the fact that the multi-body D decay-amplitude is potentially 
different at any point within the decay phase space, because of the contribution of intermediate resonances. It is shown in 
Ref.~\refcite{AS} how the rate equations for the two-body ADS method should be modified for use with multi-body final states. In the 
case of the $\rm B^{-}$ rates, for some inclusive final state $f$, Eq.~(\ref{eq:dis1}) becomes: 
\begin{equation}
\Gamma ({\rm B^{-}} \to (\bar{f})_{\rm D}{\rm K^{-}}) \propto \bar{A}_{f}^2 + r_B^{2}A_{f}^{2} + 2r_BR_{f}A_{f}\bar{A}_{f}\cos \left( 
\delta_B + \delta_D^{f} - \gamma \right), \label{eq:dis2}
\end{equation}
\noindent where $R_{f}$, the coherence factor, and $\delta_{D}^{f}$, the average strong phase difference, are defined as:
\begin{eqnarray}
A_{\rm f}^{2} & = & \int \vert A_{\rm D^{0}}(\mathbf x) \vert^{2}~d\mathbf x, 
~~~~\bar{A}_{\rm f}^{2} =  \int \vert A_{\rm \bar{D}^{0}}(\mathbf x) \vert^{2}~d\mathbf x, \\
R_{f}e^{i\delta_{\rm D}^{f}} & = & \frac{\int \vert A_{\rm D^{0}}(\mathbf x) \vert \, \vert A_{\rm \bar{D}^{0}}(\mathbf x) \vert \, 
e^{i\zeta(\mathbf x)}~d\mathbf x}{A_{\rm f} \, \bar{A}_{\rm f}} ~~~~\{ R_{f} \in \mathbb{R}~\vert~ 0 \leq R_{f} \leq 1 
\},~~\label{eq:Rf}
\end{eqnarray}
\noindent with $\mathbf x$ representing a point in multi-body phase space and $\zeta(\mathbf x)$ the corresponding strong phase 
difference. 

\section{Determining $R_{f}$ and $\delta_{D}^{f}$ at CLEO-c}
Through exploiting the fact that meson pairs produced via quarkonium resonances at $e^{+}e^{-}$ machines are in quantum entangled 
states, it is possible to obtain observables that are dependent on parameters associated with multi-body decays. In particular, it has 
been shown in Ref.~\refcite{AS} that, double-tagged $\rm D^{0}\bar{D}^{0}$ rates measured at the $\psi(3770)$ provide sensitivity to 
both the coherence factor, $R_{f}$, and the average strong phase difference, $\delta_{D}^{f}$. Starting with the anti-symmetric 
wavefunction\cite{JR} of the $\psi(3770)$ and then calculating the matrix element for the general case of two inclusive final states, 
$F$ and $G$, the double-tagged rate is found to be proportional to:
\begin{equation}
\Gamma(F|G) \propto A_{F}^{2} \, \bar{A}_{G}^{2} \, + \, \bar{A}_{F}^{2}\, A_{G}^{2} - \, 2 \, R_{F} \, R_{G} \, A_{F} \, \bar{A}_{F} 
\, A_{G} \, \bar{A}_{G} \, \cos(\delta_{D}^{F} - \delta_{D}^{G}).
\end{equation}
\noindent From this, one finds three separate cases of interest for accessing both the coherence factor and the average strong phase 
difference. These are summarised in Table~\ref{table1} below, in the instance of $F = \rm K\pi\pi\pi$.

\begin{table}[ph]
\tbl{Double-tagged rates of interest and their dependence on the coherence factor, $R_{K3\pi}$, and the average strong phase 
difference, $\delta^{K3\pi}_{D}$. The background subtracted yields from the $818~\rm{pb}^{-1}$ data sample are shown along with the 
corresponding result for each measurement.}
{\footnotesize
\begin{tabular}{cr @{=} lr @{$\pm$} l}
\toprule
$\boldsymbol{\rm K^{\pm}\pi^{\mp}\pi^{+}\pi^{-}}$ {\bf vs.} & \multicolumn{2}{c}{\bf Measurement} & 
\multicolumn{2}{c}{$\boldsymbol{\rm 818~pb^{-1}~Yield}$}\\ 
\toprule
$\rm K^{\pm}\pi^{\mp}\pi^{+}\pi^{-}$ &  $(R_{K3\pi})^{2}$~ & ~~~$0.00 \pm 0.16 \pm 0.07$ & ~~30~ & ~6\\[1ex]
CP-Tags &  $R_{K3\pi} \, \cos(\delta^{K3\pi}_{D})$~ & ~$-0.60 \pm 0.19 \pm 0.24$ & ~~2,183~ & ~47\\[1ex]
$\rm K^{\pm}\pi^{\mp}$ & $R_{K3\pi} \, \cos(\delta^{K\pi}_{D}-\delta^{K3\pi}_{D})$~ & ~$-0.20 \pm 0.23 \pm 0.09$ & ~~38~ & ~6\\[1ex]
\hline
\end{tabular}\label{table1} }
\vspace*{-18pt}
\end{table}

\subsection{Event Selection}
At present, only double-tagged samples for the determination of $R_{K3\pi}$ and $\delta^{K3\pi}_{D}$ have been analysed using CLEO-c's 
complete $\psi(3770)$ dataset, corresponding to an integrated luminosity of $\rm 818~pb^{-1}$. To maximise statistics, a total of nine 
distinct CP tags are reconstructed against $\rm K^{\pm}\pi^{\mp}\pi^{+}\pi^{-}$: $\rm K^{+}K^{-}$, $\pi^{+}\pi^{-}$, $\rm 
K_{s}^{0}\pi^{0}$, $\rm K_{s}^{0}\omega/\eta(\pi^{+}\pi^{-}\pi^{0})$, $\rm K_{s}^{0}\pi^{0}\pi^{0}$, $\rm K_{s}^{0}\phi$, $\rm 
K_{s}^{0}\eta(\gamma\gamma)$ and $\rm K_{s}^{0}\eta'(\pi^{+}\pi^{-}\eta)$. Backgrounds within these CP-tagged samples are typically 
low; in the range of $\sim 1$ to $7\%$. The flat contribution to this background is assessed from sidebands within the 
beam-constrained mass distribution for each selection, whilst peaking contributions are determined from Monte Carlo. Depending on the 
final state, the selection efficiency ranges from $\sim 4$ to $30\%$. The background subtracted yields obtained are quoted in 
Table~\ref{table1}.

\subsection{Preliminary Results}
\begin{figure}[ht]

\centerline{\epsfxsize=9cm\epsfbox{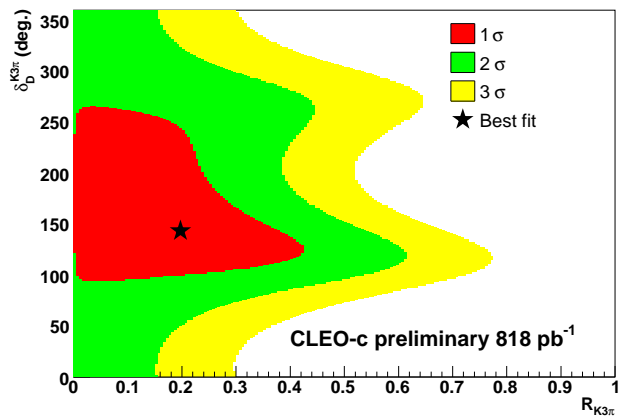}}
\vspace{-18pt} 
\caption{The resulting limits on $R_{K3\pi}$ and $\delta^{K3\pi}_{D}$ at 1, 2 and $3\sigma$ levels. \label{fig:paramspace}}
\end{figure}

From the background subtracted yields determined, central values have been calculated for $R_{K3\pi}\cos(\delta^{K3\pi}_{D})$ for each 
of the 9 CP-tags; for $R_{K3\pi}\cos(\delta^{K\pi}_{D} - \delta^{K3\pi}_{D})$ using the $\rm K^{\pm}\pi^{\mp}\pi^{+}\pi^{-}$ vs. $\rm 
K^{\pm}\pi^{\mp}$ sample; and for $(R_{K3\pi})^{2}$ using the observed number of $\rm K^{\pm}\pi^{\mp}\pi^{+}\pi^{-}$ vs. $\rm 
K^{\pm}\pi^{\mp}\pi^{+}\pi^{-}$ events. In addition, the results of the 9 separate CP-tags are used to form a combined result for 
$R_{K3\pi}\cos(\delta^{K3\pi}_{D})$, taking full account of correlations between systematic uncertainties. The preliminary results are 
quoted in the $2^{\rm nd}$ column of Table~\ref{table1}, where the first error is statistical and the second systematic. The resulting 
constraints on the parameters $R_{K3\pi}$ and $\delta^{K3\pi}_{D}$ from these measurements are shown in Fig.~\ref{fig:paramspace}.
\vspace{-12pt}
\vspace{-12pt}\noindent It is apparent, from Fig.~\ref{fig:paramspace}, that the coherence across all phase space is low, reflecting 
the fact that many out of phase resonances contribute to the $\rm K\pi\pi\pi$ final state. An inclusive analysis of this final state 
with the ADS analysis will therefore have low sensitivity to the angle $\gamma$, although the structure of Eq.~(\ref{eq:dis2}) makes 
it clear that such an analysis will allow for a determination of the amplitude ratio $r_{B}$, which is a very important auxiliary 
parameter in the $\gamma$ measurement.

\end{document}